\documentclass[showpacs,twocolumn,prl]{revtex4}
\usepackage{graphicx}

\begin{document}

\title{Energy Flow Puzzle of Soliton Ratchets}
\author{S. Denisov, S. Flach, and A. Gorbach}
\affiliation{\it Max-Planck-Institut f\"ur Physik komplexer Systeme \\
N\"othnitzer Str. 38, 01187 Dresden, Germany}

\date{\today}

\begin{abstract}
We study the mechanism of directed energy transport for soliton
ratchets. The energy flow appears due to
the progressive motion of a soliton (kink) which
is an energy carrier. However, the energy current 
formed by internal system deformations (the total field momentum) is zero.
We solve the underlying puzzle by showing 
that the energy flow is realized via 
an {\it inhomogeneous} energy exchange between the system and the
external ac driving. 
Internal kink modes are unambiguously shown to be crucial
for that transport process to take place.
We also discuss effects of spatial discretization and combination
of ac and dc external drivings.
\end{abstract}

\pacs{ 05.45.Yv, 05.60.Cd}
\maketitle

A transport mechanism of potential relevance in various 
areas of physics, chemistry and biology is based on the 
ratchet effect
\cite{1}, i.e. the generation of directed currents
by zero-mean external perturbations.
It is based on the breaking of relevant space-time symmetries of 
the underlying
system evolution equations \cite{2}. 
A paradigmatic model corresponds to a
classical particle moving in a spatially periodic potential under
the influence of zero-mean fluctuations \cite{1}, where the energy current
is directly connected with the mean particle momentum and 
the corresponding kinetic energy of the particle. 
For spatially extended systems, e.g. an annular Josephson junction, which is
described
by a partial differential equation (PDE), 
the ratchet phenomenon manifests as a
unidirectional motion of a collective kink excitation (soliton) 
\cite{3,4,5,6,molina,7}.
Here the unambiguous {\it ab initio} definition of a current 
may become a much more complicated task, because the kink 
excites other modes in the system during its motion, which may contribute
to an energy current as well.

A well-known model in the field of soliton ratchets is the
driven-damped sine-Gordon equation \cite{8}, which is also used for
modelling
the abovementioned annular Josephson junction \cite{7}:
\begin{eqnarray}
\varphi_{tt}-\varphi_{xx}=-\alpha\varphi_{t}-
\sin \varphi+E(t) \; ,
\label{1}
\end{eqnarray}
where 
$E(t)$ is a zero-mean time-periodic
driving force, $E(t+T)= E(t)$, $\int_0^T E(t) dt=0$. 
We impose the kink-bearing periodic
boundary condition:
\begin{eqnarray}
\varphi(x+L,t)=\varphi(x,t)+Q\;,\;\varphi_t(x+L,t)=\varphi_t(x,t)\;,
\label{2}
\end{eqnarray}
where $Q=2\pi m$ is the topological charge with integer $m=1,2,...$, 
and $L$ is the system size.

Let us consider the easiest case $m=1$, i.e. the presence of
one kink in the system $Q=2\pi$. The kink velocity $V$ is
defined e.g. as \cite{4,5}:
\begin{eqnarray}
V(t)=\frac{1}{Q}\int^{L}_{0} x \varphi_{tx}dx\;.
\label{3}
\end{eqnarray}
In the case of a soliton ratchet the mean value of $V(t)$ will be nonzero.
Since the kink carries some energy, one expects a mean nonzero energy 
current as well.
Recently it has been proposed to observe this
directed energy transport using the definition of the {\it internal energy
current} $J^{I}$ and its density $j^{I}$ \cite{5}:
\begin{eqnarray}
J^{I}(t)=\int_{0}^{L} j^{I} dx \;,\;
j^{I}(x,t) = - \varphi_{x}
\varphi_{t}\;.
\label{4}
\end{eqnarray}
$J^I$ is also known as the {\it total momentum} of the system \cite{8,9,10}.

Choosing either Eq.(\ref{3}) or Eq.(\ref{4}), 
the symmetry analysis provides identical necessary conditions 
for the appearance of a ratchet effect \cite{5}.
If the ac driving $E(t)$ 
possesses a shift symmetry,
\begin{eqnarray}
E(t)= -E(t+T/2)\;,
\label{5}
\end{eqnarray}
then the combined symmetry transformation
\begin{eqnarray}
x \rightarrow -x, ~~ \varphi \rightarrow -\varphi + Q, ~~ t
\rightarrow t + \frac{T}{2}
\label{6}
\end{eqnarray}
leaves the equation (\ref{1}) invariant and changes the 
sign of $V$ and $J^{I}$. Consequently, if Eq.(\ref{1}) allows
for only one attractor solution, both quantities will have average
value zero.
Violating (\ref{5}) we loose the symmetry (\ref{6}) and 
may expect nonzero values for the mean values of $V$ and $J^{I}$.
This can be done e.g. by the choice \cite{5,6,7}:
\begin{eqnarray}
E(t)= E_{1}\cos(\omega t)+ E_{2}\cos(2 \omega t + \Theta)\;,\; 
\omega=\frac{2\pi}{T}\;.
\label{7}
\end{eqnarray}

The soliton ratchet effect has been observed in terms of the 
mean soliton velocity $V$ both numerically \cite{3,4,5}
and experimentally in an annular Josephson junction \cite{7}.
However, by differentiating  Eq.(\ref{4}) and using Eq.(\ref{1}) together 
with the boundary condition (\ref{2}) it follows that \cite{9,11}:
\begin{eqnarray}
J^{I}_{t}(t)= -\alpha J^{I}(t)-Q E(t)\;.
\label{8}
\end{eqnarray}
Thus, for any ac driving force $E(t)$ with zero mean the 
time-averaged value of the total momentum, $J^{I}= \lim_{t\rightarrow \infty}
\frac{1}{t}\int_{0}^{t}J^{I}(\tau)d \tau$, is zero.
Suppose that a kink moves with some non-zero mean velocity $V$. 
The kink carries  also some non-zero energy.
Hence the progressive kink motion should 
lead to the appearance of an energy flux in  the system.
Thus the system supports a {\it non-zero} averaged energy flow  
and a {\it vanishing} averaged
internal energy current (total system momentum). We will 
solve the puzzle in this Letter.

The PDE 
(\ref{1}) corresponds to the energy density $\rho$
\begin{eqnarray}
\rho[\varphi(x,t)] \equiv \rho(x,t)=\frac{1}{2}(\varphi_{t}^{2} +
\varphi_{x}^{2})+1-\cos (\varphi)\;.
\label{9}
\end{eqnarray}
Using
Eq.(\ref{1}) we obtain:
\begin{eqnarray}
%\nonumber
\rho_{t}=
%\frac{\partial}{\partial x} [\varphi_{x}
%\varphi_{t}]-\alpha
%\varphi_{t}^{2}+E(t)\varphi_{t}= \\
-j^{I}_{x}-\alpha
\varphi_{t}^{2}+E(t)\varphi_{t}\;.
\label{10}
\end{eqnarray}
The last two terms  describe energy losses through
dissipation and  the energy exchange between the system and external 
driving $E(t)$. 
However they also contain a derivative $j^E_x$ of an additional
current density, as we will show below.
A central result of this work is that
these terms describe a new 
{\it exchange} energy  current $J^{E}(t)$. This exchange current
corresponds to an
additional energy transmission channel 
provided by a spatially {\it  inhomogenous} 
energy exchange between the system,
the external ac driving and the dissipation. Thus, the complete 
current balance equation for the full current $J$ reads:
\begin{eqnarray}
J(t)= {J}^{I}(t)+{J}^{E}(t)\;.
\label{11}
\end{eqnarray}

We consider a large system size,
i.e. $ L \gg L_{k}$, where
$L_{k}$ is the kink localization length. Far from the kink center
the field is oscillating in time while being homogeneous in space.
We also assume that the kink travels only over distances 
$L_{p} \ll L$
during one period of the ac driving.
Because of the external ac force the field evolution is uniquely locked
to the driver (as it was the case
for all previous considerations \cite{3,4,5,6,7}, i.e. assuming that
the system allows only for one attractor). 
Thus we observe a moving
localized excitation (kink) which propagates on a background
formed by a spatially homogenous groundstate. 
It is convenient to
separate $\varphi(x,t)$ into a localized kink part,
$\varphi^{k}(x,t)$, where $\varphi^{k}(x\rightarrow 0,t)=
0;~\varphi^{k}(x\rightarrow L,t)= 2\pi$, and a background ({\it vacuum})
part $\varphi^{v}(t)$ which depends only on time \cite{9}:
\begin{eqnarray}
\varphi(x,t)= \varphi^{k}(x,t)+\varphi^{v}(t)\;.
\label{12}
\end{eqnarray}
The vacuum part alone must satisfy Eq.(\ref{1}). 
Because it is also a solution of the system in the absence of 
a kink when $Q=0$, it can not contribute to any energy
transport \cite{5}.

On the attractor the dynamics of the system (\ref{1}) is given by
%realized on a winding attractor:
\begin{eqnarray}
\varphi^k(x,t+T)=\varphi^k(x-VT,t)\;,
\;
\varphi^{v}(t+T)=\varphi^{v}(t)\;,
\label{13}
\end{eqnarray}
where $T=2\pi/\omega$ and 
$V$ is the averaged kink velocity, 
$V=\langle V(t)\rangle_{T}=\frac{1}{T}\int_{0}^{T} V(t) dt$.
Note that all integral system characteristics, such as the 
total energy of the system, the kink velocity, and
energy currents from Eq.(\ref{11}) are periodic functions of time 
with the period $T$.

Let us compute the full energy current $J$ produced 
by a moving kink.
We use 
the periodicity of the total system energy $W(t)=\int_{0}^{L}
\rho(x,t)dx $ in time. The amount of energy carried through a 
point $x$ between time $t$ and $t+T$ is equal to:
\begin{eqnarray}
\nonumber
\triangle w(x,t)= \int_{0}^{x} [\rho (x',t) - \rho (x',t+T)]
dx' =  \\
-\int_{0}^{x} dx' \int_{t}^{t+T} \rho_{t} (x',t') dt'\;.
\label{14}
\end{eqnarray}
Here we assume that at $x=0$ we have a homogenous vacuum state
during the full cycle of the ac driving, thus 
there is no energy current through this point.
Due to the presence  of ac driving the system is not time
homogeneous and $\triangle w(x,t)$ in Eq.(\ref{14}) depends on $t$.
With (\ref{2}),(\ref{13}) it follows $\triangle w(x,t+T)=\triangle w(x-VT,t)$
and $\triangle w(x+L,t)=\triangle w(x,t)$ and
\begin{eqnarray}
J=\frac{1}{T} \int_0^T dt \int_0^L dx j(x,t) \;,
\;
j(x,t)=\frac{1}{T} \triangle w(x,t) \;.
\label{totalcurrent}
\end{eqnarray}
Using Eqs.(\ref{2}),(\ref{13}) and  the multiplicative integration
rule \cite{12} 
we obtain from (\ref{14}),(\ref{totalcurrent}):
\begin{eqnarray}
J=V  \langle \int_{0}^{L} \rho[\varphi(x,t)] dx -
L  \rho[\varphi^{v}(t)] \rangle_{T} 
\label{16}
\end{eqnarray}
with $\langle ... \rangle = 1/T \int_0^T ... dt$.
The quantity being averaged in the rhs of (\ref{16}) is
the  difference  between the
energy of system with and without a kink, or
simply 
the kink energy $W^{k}=W_{Q}(t)-W_{0}(t)$, 
so that the mean total current, Eq.(\ref{16})
generated by the moving kink reads
\begin{eqnarray}
J=VW^{k}\;.
\label{17}
\end{eqnarray}
It follows that the total energy 
current has the same symmetry properties as the kink velocity (\ref{3}).
This result proves the initial intuitive guess that a 
moving kink indeed generates a non-zero energy current in the system.
\begin{figure}[t]
\includegraphics[angle=-90, width=0.7\linewidth]{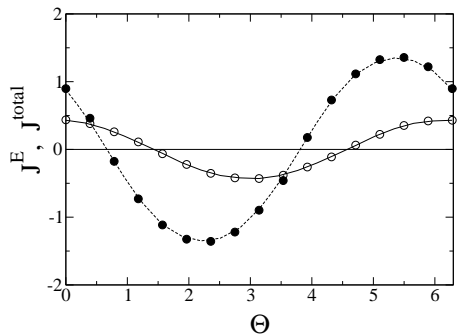}
\caption{ The dependence of the mean exchange current $J^{E}$ 
on $\Theta$
for $\alpha=0.2$ (solid line) and $\alpha=0.05$ (dashed
line). Circles correspond to the numerical results for
$J^{total}$. Other parameters: $E_1=E_2=0.2$, $\omega=0.1$, $Q=2\pi$, $L=500$.
\label{fig1}}
\end{figure}

Because for any
ac driving $E(t)$ with zero mean the averaged internal current $J^{I}$
is equal to zero, we arrive at
the following energy current balance:
\begin{eqnarray}
J=J^{E}, ~~~ J^{I}=0\;.
\label{18}
\end{eqnarray}

In order to obtain an expression for the exchange current density,
we use (\ref{10}) and (\ref{totalcurrent}) and arrive at
\begin{eqnarray}
j^{E}(x,t)= -\frac{1}{T} \int_{t}^{t+T} dt' \int_{0}^{x} dx'
\phi(x',t') \;, 
\label{19}
\\
\phi(x,t)=\phi[ \varphi ]= \alpha \varphi_{t}^{2} -
E(t)\varphi_{t}\;,
\label{20}
\\
\label{fullexchangecurrent}
J^E = \frac{1}{T} \int_0^T dt \int_0^L dx j^E(x,t)\;.
\end{eqnarray}
Evidently the exchange current
$J^{E}$ has the same symmetry as $V$, $J$ and $J^{I}$.

We solved the equation (\ref{1}) numerically \cite{13} 
in order to test the current balance (\ref{18}).
A crucial parameter is the mesh size $h$ used for the spatial
discretization of the equation (\ref{1}). For any finite value of $h$
the internal current $J^{I}$ is nonzero, as also obtained in \cite{5}. 
However it scales according to $J^{I} \sim h^{2}$ for $h \leq 0.1$
and vanishes in 
the continuum limit $h \rightarrow 0$ in full accord with (\ref{18}).
We chose $h=0.1$ here, for which 
the values of $J$ and $J^E$ are determined
with an error of less than $3\%$, while $J^{I}/J \sim 0.003$
(for $h=0.32$ used in \cite{5} the latter ratio increases to 0.02 for
$\alpha=0.2$ and 0.18 for $\alpha=0.05$).
The results of numerical calculations of  $J$,
Eq.(\ref{16}), and $J^{E}$, Eq. (\ref{fullexchangecurrent}),  
are shown in Fig.\ref{fig1}.
We obtain very good agreement with (\ref{18}).
Note that $L=500$, $L_{k}\approx 10$ and $L_{p} \approx 30$.
According to the collective coordinate theories \cite{6,molina}
a nonzero average kink velocity implies the excitation of
internal shape modes on the kink. Thus the appearance of a nonzero
exchange current is another manifestation of the excitation
of internal kink modes for the case of the considered soliton
ratchet.

\begin{figure}[t]
\includegraphics[width=1.\linewidth,angle=0]{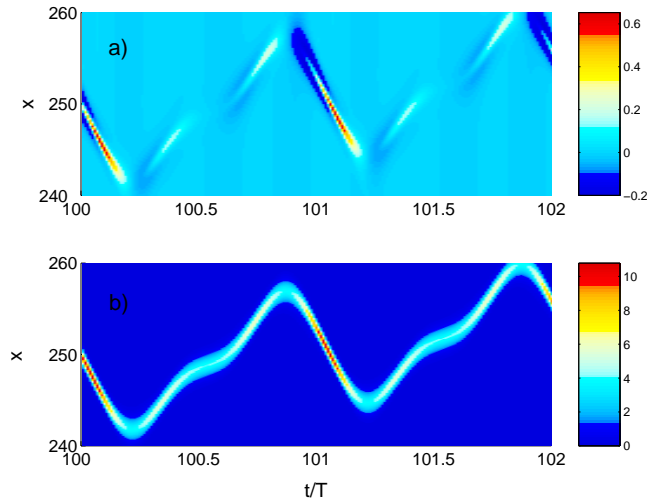}
\caption{ Space-time evolution of the soliton ratchet
for $\alpha=0.2$ and other parameters as in Fig.\ref{fig1}.
(a) Contour plot of the function
$\phi(x,t)$, Eq. (\ref{20}). 
(b) Contour plot of the energy density (\ref{9}).
\label{fig2}}
\end{figure}

In Fig.\ref{fig2}(a) we plot the space-time evolution of the function
$\phi(x,t)$ as defined in (\ref{20}), which describes the energy exchange
between the field $\varphi$, the external ac drive $E(t)$ and the
friction term. 
The energy is exchanged and transported
in a cyclic way:
first the kink absorbs energy in its rare tail, then it releases
energy in its front, then it absorbs energy in its front and
finally releases energy in the rare tail.
In Fig.\ref{fig2}(b) we plot the space-time evolution
of the energy density $\rho(x,t)$ (\ref{9}). The excitation of
internal shape modes on the kink is clearly observed - the kink
is much more compressed when moving opposite to its average propagation
direction as compared to the times when it moves in the same direction.
 
It is very instructive to apply 
our approach based on the 
generalized current balance equation (\ref{11}) to
the well-studied case of a constant force, $E=const$ \cite{8}.
Due to the time homogeneity of the system
$\varphi(x,t)=\varphi(\xi)\;,\;\xi=x-Vt$.
Still we deal with the motion of a spatially localized kink
with a {\it finite extension}. 
The internal current $J^{I}$ can be
evaluated using Eq.(\ref{8}):
\begin{eqnarray}
J^{I}=-\frac{EQ}{\alpha}\;.
\label{22}
\end{eqnarray}
The internal current density 
\begin{equation}
j^I(\xi)=V\varphi_x^2|_{x\equiv \xi}\;.
\label{internalcurrentdensityconstantfield}
\end{equation}
Similarly the exchange current density can be obtained as
\begin{equation}
j^E(\xi)=\int_0^{\xi} dx (\alpha V^2\varphi_{x}^2 + VE\varphi_x)\;.
\label{exchangecurrentdensityconstantfield}
\end{equation}
Due to the time homogeneity the kink energy $W_k$  is now independent
on time, and the total current can be computed using (\ref{17}).
\begin{figure}[t]
\includegraphics[angle=-90, width=0.8\linewidth]{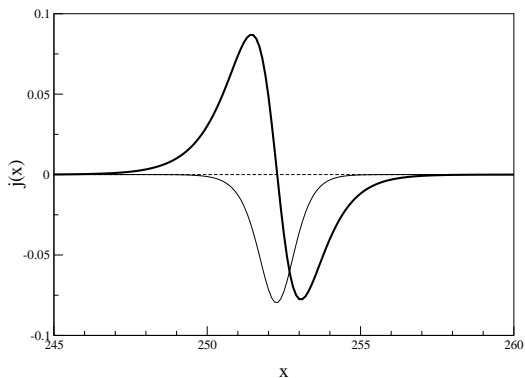}
\caption{ The densities of the scaled internal current, $0.02 j^{I}(x)$ (thin
line), and the exchange current, $j^{E}(x)$ (bold line), for the
 constant force case $E=0.2$.
The other parameters are $Q=2\pi$, $L=500$ and
$\alpha=0.2$.
\label{fig3}}
\end{figure}
We plot the current densities in Fig.\ref{fig3}. 
The internal current density is single peaked, with the peak position
corresponding to the position of the kink center.
The
exchange current density $j^E \neq 0$, which may come as a surprise,
since in this case the kink moves {\it without} exciting internal
shape modes. Yet the kink is a spatially extended object, and it is
this property which leads to a nonzero exchange current density.
We also computed the kink velocity, the kink energy, the total
current and the total internal and exchange currents for the
case $E=0.2$ and $\alpha=0.2$, $L=670$ and $h=0.045$. 
The kink velocity $V=-0.6201$,
the kink energy $W_k=10.0813$ and thus the total current $J=-6.251$.
The internal current $J^I=-6.282$, which is off the exact result
$J^I=-2\pi$ (\ref{22}) by an error $\delta=0.001$. The total exchange current
is obtained as $J^E=0.0321$. The current balance equation 
(\ref{11}) is satisfied within the same small error $\delta$, which
is an order of magnitude smaller than the computed value for $J^E$.
Together with a series of further tests, including the analysis of
the dependence of the results on $L$ and $h$, we conclude that 
{\it the exchange current is nonzero for a constant field $E$}.
It follows from the mismatch between the total current $J=V W_k$
and the internal current $J^I$. Since no internal modes are
excited on the kink in this case, the appearance of an exchange
current (or its density) is simply linked to the finite spatial extent
of the kink, and the corresponding spatially (and thus temporally as well)
inhomogeneous energy exchange between the field $\varphi$ and the
external field $E$ and the friction term, similar to the soliton ratchet.
Notably the exchange current for a constant field $E$
will vanish in the Hamiltonian
limit $E=\alpha=0$ and also in the overdamped limit $\alpha \gg 1$.

Let us combine both cases from above, i.e. 
both  a constant 
and an ac components of the driving $E$. 
A careful tuning of the parameters leads to an exact cancellation
of both force components and $V=0$. 
At that point
$J=0$ (\ref{17}). At the
same time, according to Eq. (\ref{22}), the internal current is $J^{I} =
-QE^{stop}/\alpha$ where $E^{stop}$ is the dc component of $E$.
This implies an
exact
balance between the two currents,
\begin{eqnarray}
J^{total}=0,~~ J^{E} = -J^{I}\;.
\label{25}
\end{eqnarray}
It is also possible to have a situation where
the
sign of the total momentum of the system $J^{I}$, Eq.(\ref{4}), is 
{\it opposite} to the sign of the kink velocity $V$,
and the internal current is pumping energy against the kink motion.

We solved the soliton ratchet puzzle of nonvanishing kink velocities
and vanishing total momentum (see also \cite{14} and \cite{10}).
We identified a new energy pathway
which is entirely mediated by the spatial and temporal inhomogeneity
of the system mediated by the ac driving and the spatial extension
of the kink. More precisely, internal kink modes are crucial
in order to sustain this energy exchange current for the soliton ratchet.
The reason for the appearance of this new exchange current is
provided by the finite spatial extent of the kink.
Even for the case of a constant external field, the exchange current
is found to be nonzero.

Discretizing space will  
additionally change the ratio between the various currents.
For the ac driving case of a soliton ratchet, corrections to
the internal current invalidate (\ref{8}) 
so that
$J^{I}$  also contributes to the total energy flow \cite{5}
even for a zero mean drive.
While our results are important for the general case of
spatially extended fields coupled to external driving fields or
simply other degrees of freedom, they also apply directly
to the soliton ratchet in an annular Josephson junction \cite{7}.

\begin{acknowledgements}
We thank N.R. Quintero for drawing our attention to \cite{11}, and 
Y. Zolotaryuk and V. Fleurov for helpful discussions.
\end{acknowledgements}


\begin{thebibliography}{99}
\bibitem{13} We deal with a
discrete version of the system in Eq.(\ref{1}), 
$x \rightarrow h n$, $n=0,1,2,...,N$,
$\varphi(x,t) \rightarrow \varphi(hn,t)=\varphi_{n}(t)$, 
and $\varphi_{xx}
\rightarrow h^{-2}(\varphi_{n+1}-2\varphi_{n}+\varphi_{n})$. 
For the integration
of the corresponding set of coupled ordinary differential
equations we used a standard
fourth-order Runge-Kutta method. 
The deviation of any computed number from its continuum limit value 
was monitored as a function of $h$.
\end{thebibliography}
\end{document}